\documentclass[10pt]{article}
\usepackage{epsfig}
\usepackage{lineno}
\usepackage{cite}

\newcommand{\be}{\begin{equation}}
\newcommand{\ee}{\end{equation}}
\newcommand{\bea}{\begin{eqnarray}}
\newcommand{\eea}{\end{eqnarray}}

\begin{document}
\sloppy

\title{ \vspace{1cm} Spectroscopy of $^{19}$Ne for the thermonuclear $^{15}$O($\alpha,\gamma$)$^{19}$Ne and $^{18}$F($p,\alpha$)$^{15}$O reaction rates}
\author{A. Parikh$^{1,2,} \footnote{xrayburst@gmail.com}$, A.M. Laird$^{3}$, N. de S\'er\'eville$^{4}$, K. Wimmer$^{5}$, T. Faestermann$^{6}$, R. Hertenberger$^{7}$, \\D. Seiler$^{6}$, H.-F. Wirth$^{7}$, P. Adsley$^{3}$, B.R. Fulton$^{3}$, F. Hammache$^{4}$, J. Kiener$^{8}$, I. Stefan$^{4}$\\
\\
\small $^1$Departament de F\'isica i Enginyeria Nuclear, \\ \small Universitat
Polit\`ecnica de Catalunya, E-08036 Barcelona, Spain\\
\small $^2$Institut d'Estudis Espacials de Catalunya (IEEC), E-08034 Barcelona, Spain\\
\small $^3$Department of Physics, University of York, York YO10 5DD, United Kingdom\\
\small $^4$Institut de Physique Nucl\'eaire d'Orsay, UMR8608, IN2P3-CNRS, Universit\'e Paris Sud 11, 91406 Orsay, France\\
\small $^5$Department of Physics, Central Michigan University, Mount Pleasant, Michigan 48859, USA\\
\small $^6$Physik Department E12, Technische Universit\"at M\"unchen, D-85748 Garching, Germany\\
\small $^7$Fakult\"at f\"ur Physik, Ludwig-Maximilians-Universit\"at M\"unchen, D-85748 Garching, Germany\\
\small $^8$CSNSM, IN2P3-CNRS, Universit\'e Paris Sud 11, 91405 Orsay, France}
\maketitle

\begin{abstract} 

Uncertainties in the thermonuclear rates of the $^{15}$O($\alpha,\gamma$)$^{19}$Ne and $^{18}$F($p,\alpha$)$^{15}$O reactions affect model predictions of light curves from type I X-ray bursts and the amount of the observable radioisotope $^{18}$F produced in classical novae, respectively.   To address these uncertainties, we have studied the nuclear structure of $^{19}$Ne over $E_{x} = 4.0 - 5.1$ MeV and $6.1 - 7.3$ MeV using the $^{19}$F($^{3}$He,t)$^{19}$Ne reaction.  We find the $J^{\pi}$ values of the 4.14 and 4.20 MeV levels to be consistent with $9/2^{-}$ and $7/2^{-}$ respectively, in contrast to previous assumptions.  We confirm the recently observed triplet of states around 6.4 MeV, and find evidence that the state at 6.29 MeV, just below the proton threshold, is either broad or a doublet.  Our data also suggest that predicted but yet unobserved levels may exist near the 6.86 MeV state.  Higher resolution experiments are urgently needed to further clarify the structure of $^{19}$Ne around the proton threshold before a reliable $^{18}$F($p,\alpha$)$^{15}$O rate for nova models can be determined. 

\end{abstract}

{\bf PACS numbers:} 21.10.-k, 27.20.+n, 25.55.Kr, 26.50.+x 

\twocolumn

\section{Introduction}

Explosive thermonuclear burning of fuel accreted from a companion star onto the surface of a white dwarf or neutron star gives rise to the astrophysical phenomena known as classical novae and type I X-ray bursts (XRBs), respectively.   Within the galaxy, several hundred novae have been discovered to date along with $\approx100$ systems exhibiting XRBs.  A typical nova explosion ejects $\approx 10^{-4} - 10^{-5}$ $M_{\odot}$ of material into the interstellar medium.  Through spectroscopic analysis the chemical composition of this ejected material can, in principle, be compared to nova model predictions.  The composition of the ejecta, if any, from XRBs is still unclear.  Nonetheless, XRB models suggest that the energy released during XRBs as well as the time structure of the light curves may serve to probe details of these explosions.  In both cases, however, model predictions of observables depend upon various factors, including the assumed accretion rate, the composition of the accreted material, the mass and composition of the underlying compact object, and the nuclear reaction rates adopted.  For reviews, see, e.g., Ref. \cite{Bod08} for classical novae and Refs.\cite{Lew93,Str06,Par13} for XRBs.            

The $^{18}$F($p,\alpha$)$^{15}$O and $^{15}$O($\alpha,\gamma$)$^{19}$Ne reaction rates have significant and demonstrated impact on the predicted abundance of radioactive $^{18}$F in novae and the profiles of XRB light curves, respectively.   The former is of interest as the decay of $^{18}$F (t$_{1/2}$ = 110 min) could provide a detectable, prompt $\gamma$-ray signature of a nova at and below 511 keV through electron-positron annihilation\cite{Lei87}.  The distance at which this emission may be detected by a suitable instrument aboard a satellite, however, depends upon the amount of $^{18}$F produced during the explosion, and the  $^{18}$F($p,\alpha$)$^{15}$O reaction is the principal means of destruction of $^{18}$F at temperatures encountered in novae.  As to the  $^{15}$O($\alpha,\gamma$)$^{19}$Ne reaction, XRB models have revealed that the adopted rate has a strong impact not only on the predicted peak luminosity of a burst\cite{Dav11}, but also on whether bursting behaviour or stable burning is predicted\cite{Fis06,Kee14}.  For a recent review of the impact of nuclear physics uncertainties on predicted yields and light curves from novae and XRBs, see Ref.\cite{Par14a}.  

Direct measurement of the $^{18}$F($p,\alpha$)$^{15}$O and $^{15}$O($\alpha,\gamma$)$^{19}$Ne reactions at all relevant energies above the proton and alpha particle thresholds in $^{19}$Ne (S$_{p}$ = 6410 keV, S$_{\alpha}$ = 3528 keV\cite{Aud12}) is not yet possible due to the lack of sufficiently intense radioactive $^{18}$F and $^{15}$O beams (see Ref.\cite{Bee11} for progress with the former case).  Indirect methods must therefore be exploited: nuclear structure information is required for states within $E_{x}$($^{19}Ne)\approx6 - 7$ MeV and $\approx4 - 5$ MeV to estimate the relevant rates of the $^{18}$F($p,\alpha$)$^{15}$O and $^{15}$O($\alpha,\gamma$)$^{19}$Ne reactions, respectively.  In particular, excitation energies, $J^{\pi}$ values, total widths, and proton, alpha particle, and $\gamma$-ray partial widths are needed.  Despite considerable experimental and theoretical effort to better constrain these rates (see, e.g., Refs.\cite{Nes07, Ser09, Bee11, Ade12, Mou12, Lai13, Che15, Tan09, Dav11} and references within), the nuclear physics uncertainties are still sufficient to affect, for example, predictions of $^{18}$F production in novae by at least a factor of two\cite{Lai13} and predictions of peak XRB luminosities by a factor of $\approx2$\cite{Dav11}. 

In the present work we have measured the $^{19}$F($^{3}$He,t)$^{19}$Ne reaction to address several outstanding issues regarding the nuclear physics input used in current estimates of the $^{18}$F($p,\alpha$)$^{15}$O and $^{15}$O($\alpha,\gamma$)$^{19}$Ne rates in classical novae and XRBs, respectively.  These include (a) measuring, for the first time, the $J^{\pi}$ values of the $E_{x}$($^{19}$Ne) = 4.14 and 4.20 MeV levels, as well as of several other levels above the $^{15}$O+$\alpha$ threshold\cite{Til95,Tan09,Dav11}; (b) searching for several predicted\cite{Nes07}, but yet unobserved levels near the $^{18}$F+p threshold; and, (c) investigating the triplet at 6.4 MeV, as well as a possible doublet at 6.3 MeV, through a measurement with better energy resolution than our previous study\cite{Lai13}, in which these issues were first identified.  We conclude by briefly discussing additional measurements needed to improve estimates of these two rates.

\section{Experiment}

The $^{19}$F($^{3}$He,t)$^{19}$Ne reaction was measured at the Maier-Leibnitz-Laboratorium (MLL) in Garching, Germany using a 25-MeV $^{3}$He$^{2+}$ beam ($I = 200 - 500$ nA) and a quadrupole-dipole-dipole-dipole magnetic spectrograph.  More details on the method and equipment may be found in Refs.\cite{Lai13,Par14b} and references within.  Targets included 10 $\mu$g/cm$^{2}$ MgF$_{2}$ and 50 $\mu$g/cm$^{2}$ CaF$_{2}$, each deposited upon a 10 $\mu$g/cm$^{2}$ foil of enriched $^{12}$C (99.99\%).  As well, thin $^{12}$C and MgO targets were used to characterize any background.  Spectrograph apertures of either $d\Omega$ = 7.8 or 13.9 msr were used, and measurements were made at laboratory angles $\theta_{lab}$ between 10$^{\circ}$ and 50$^{\circ}$.   Tritons were identified and selected through energy loss and residual energy information from the detection system at the focal plane of the spectrograph, and triton spectra of focal-plane positions were then produced for further analysis.  Several different magnetic field settings were employed since associated $^{19}$Ne levels spanning only $\approx1$ MeV of excitation energy were observed at any one time due to the momentum bite of the spectrograph.   Tritons from ($^{3}$He,t) reactions on target contaminants such as $^{12}$C, $^{16}$O, $^{24}$Mg and $^{40}$Ca were excluded from the focal-plane detector at all settings because of the significantly different Q-values involved.

\section{Results}

\begin{figure}[h]
\begin{center}
\includegraphics[scale=0.58]{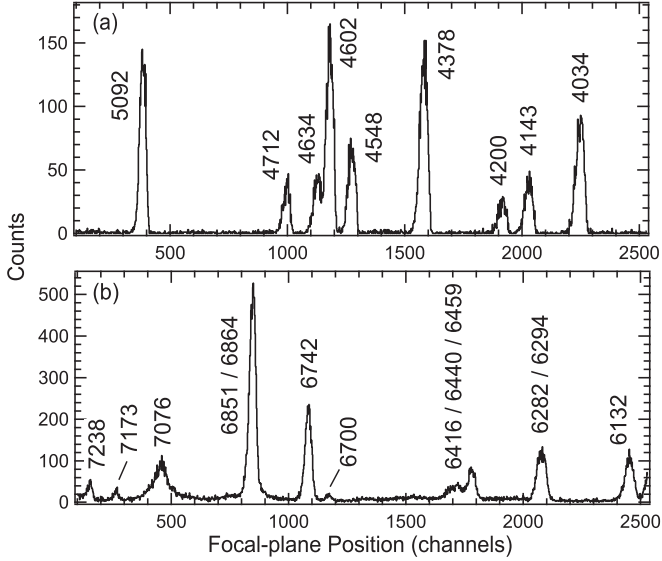}
\caption{Focal-plane position spectra of tritons from the $^{19}$F($^{3}$He,t)$^{19}$Ne reaction
at 25 MeV.  The spectra were measured at $\theta_{lab}$ = 10$^{\circ}$ using the 50 $\mu$g/cm$^{2}$ CaF$_{2}$ target (panel a), and $\theta_{lab}$ = 15$^{\circ}$ using the 10 $\mu$g/cm$^{2}$ MgF$_{2}$ target (panel b). Peaks are labeled by the corresponding $^{19}$Ne excitation energies in keV.}
\label{fig1}
\end{center}
\end{figure}

Figure \ref{fig1}(a) shows a triton position spectrum measured with the CaF$_{2}$ target over $E_{x} = 4.0 - 5.1$ MeV at $\theta_{lab}=10^{\circ}$ and $d\Omega = 13.9$ msr.  The labeled excitation energies are from Refs.\cite{Tan09,Dav11} and were not determined in the present work.   Figure \ref{fig1}(b) shows a triton spectrum measured with the MgF$_{2}$ target over $E_{x} = 6.1 - 7.3$ MeV at $\theta_{lab}=15^{\circ}$ and $d\Omega = 7.8$ msr.   The labeled excitation energies are from Refs.\cite{Lai13,Nes07}, except for the peaks designated ``6282/6294" and ``6851/6864", which are discussed below.  These spectra were analyzed using least-squares fits of multiple exponentially modified Gaussian functions (as well as Voigt functions, for broad states) with a constant or linearly-varying background.  The energy resolution was determined to be $\approx 15$ keV (panel a) and $\approx 10$ keV (panel b) FWHM from the widths of fits to isolated narrow peaks in these spectra.  The resolution of the latter spectrum improves upon that from Ref.\cite{Lai13} by a factor of $\approx1.5$ by virtue of the thinner target employed here, as well as, to a lesser extent, the narrower spectrograph aperture.

\begin{figure}[h]
\begin{center}
\includegraphics[scale=0.6]{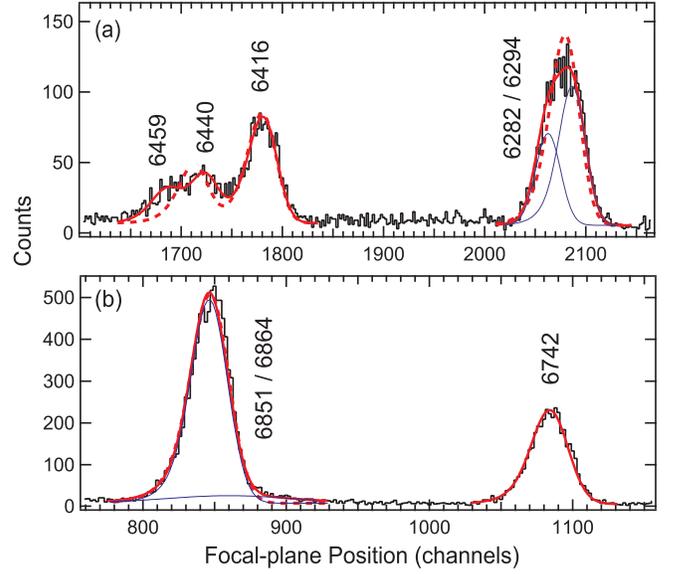}
\caption{(Color online) Expanded views of Fig. 1(b).  Overall best fits (solid red), constituent peaks (blue) and alternative, poorer fits (dashed red) are shown for levels within $E_{x}(^{19}Ne) = 6.2-6.9$ MeV.  See text for details.}
\label{fig2}
\end{center}
\end{figure}

Figure \ref{fig2} shows enlarged views of the spectrum in Fig. \ref{fig1}(b).  As shown in Fig. \ref{fig2}(a), the observed structure around 6.4 MeV is best fit with a triplet of states (solid line, two-tailed $p$-value = 0.48) as opposed to a doublet (dashed line, $p <$ 0.0001) when widths from isolated narrow states (e.g., at 6132, 6742, 7173, or 7238 keV) are adopted.  Similarly for the structure at 6.29 MeV, a doublet of states is strongly preferred (solid lines, $p$ = 0.59) over a single narrow state (dashed line, $p <$ 0.0001), as shown in Fig. \ref{fig2}(a).  This structure may be a single broad state, in which case assuming a state with a width of $\approx16$ keV provides an acceptable fit ($p$ = 0.23).  Indeed, the lack of any obvious angular distribution effects indicating the contribution of two components supports the identification of this peak as a single state, although the contributing levels may simply have similar spin and parity.  Finally, the structure at 6.86 MeV is poorly fit with a single narrow state due to the excess of counts at low excitation energies (dashed line, $p <$ 0.0001).  Assuming two or three contributing narrow states or a single, broad state does not significantly improve the fit, giving $p$-values of less than 0.03.  A reasonable fit is obtained using a single narrow state together with a broad state of width $\approx40$ keV (solid lines, $p$ = 0.21).  If we use the energies from Refs.\cite{Nes07,Lai13} for the levels at 6.13, 6.70, 6.74, 7.08, 7.17, and 7.24 MeV and internally calibrate this spectrum, we obtain energies that are consistent with those reported in Ref. \cite{Lai13} for the triplet at 6.4 MeV, an assumed single state at 6.86 MeV, and an assumed broad state at 6.29 MeV.  If we assume the structures at 6.29 and 6.86 MeV are doublets, we find the constituent levels to have energies of 6282(2) and 6295(2) keV, and 6851(4) and 6864(1) keV, respectively.  For the $^{18}$F($p,\alpha$) reaction, these would correspond to resonances at $E_{c.m.} = -128, -115,  441,$ and 454 keV.  Results consistent with the above discussion were obtained at all angles measured in the present work.

\begin{figure}[h]
\begin{center}
\includegraphics[scale=0.55]{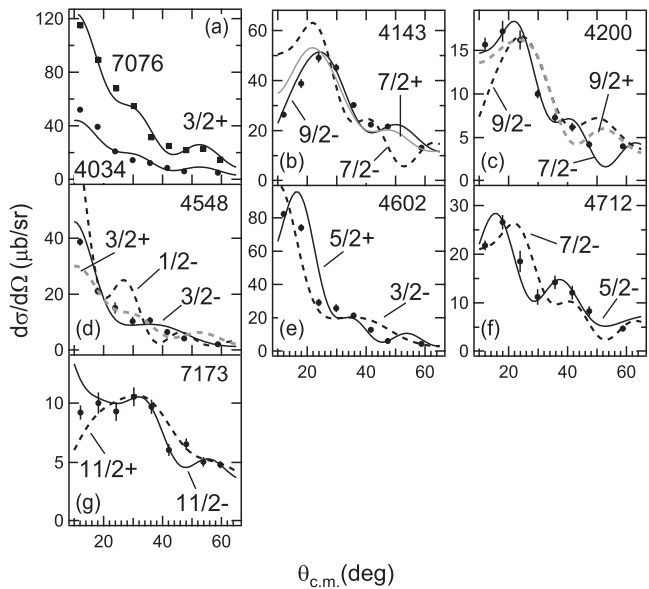}
\caption{Triton angular distributions measured with the $^{19}$F($^{3}$He,t)$^{19}$Ne reaction at 25 MeV.  Theoretical curves calculated with FRESCO\cite{FRESCO} have been fit to the data.  Best fit curves are plotted here, along with curves corresponding to previous, tentative assignments.  Excitation energies (in keV) corresponding to the data and $J^{\pi}$ values corresponding to each curve are indicated.  See text for full details.}
\label{fig3}
\end{center}
\end{figure}

Figure \ref{fig3} shows measured angular distributions for states in $^{19}$Ne populated through the $^{19}$F($^{3}$He,t)
reaction.  These differential cross sections have been fit using theoretical calculations from the finite-range, coupled-channels reaction code FRESCO~\cite{FRESCO} as discussed in Refs.\cite{Lai13,Par11}.  Acceptable fits ($p >$ 0.05) were obtained for theoretical calculations corresponding to $J^{\pi}$ values of observed states with well-known, measured $J^{\pi}$ values\cite{Til95, Nes07}, giving confidence in the predictive power of the angular distributions.  For example, fitting the $3/2^{+}$ theoretical curve to the distributions for the 4034 and 7076 keV levels, both $3/2^{+}$, yielded $p$-values of 0.39 and 0.69, respectively.  No other theoretical curve provided an acceptable fit to the 7076 keV state, while the $3/2^{-}$ curve gave the only other reasonable fit ($p$ = 0.19) to the 4034 keV state.  We discuss levels with previously unmeasured $J^{\pi}$ values\cite{Til95,Nes07} below.  Angular distributions for levels between 6.1 and 6.9 MeV (assuming the 6.29 and 6.86 MeV states to be single levels) were consistent with those reported in Ref.\cite{Lai13} and are not repeated here.  

The 4.14 MeV and 4.20 MeV levels have been tentatively assigned either $(9/2)^{-}$ and $(7/2)^{-}$ \cite{Til95} or $(7/2)^{-}$ and $(9/2)^{-}$\cite{Tan09}, respectively.  The former is best fit by the $9/2^{-}$ ($p$ = 0.99) and $7/2^{+}$ ($p$ = 0.34) curves.  The $7/2^{-}$ curve is a poor fit to the data ($p <$ 0.01) especially at low angles.  For the latter level, $7/2^{-}$ and $9/2^{+}$ both provide reasonable fits ($p$ = $0.85 - 0.99$).  The $9/2^{-}$ curve gives a significantly worse fit to the data ($p$ = 0.09), again at the lowest measured angles.  Given the established parity of these states, we assign $9/2^{-}$ and $7/2^{-}$ to the 4.14 and 4.20 MeV levels.

The 4.55, 4.60 and 4.71 MeV levels have tentative assignments of $(1/2,3/2)^{-}$, $(5/2^{+})$, and $(5/2^{-})$, respectively\cite{Til95}.  The first of these is best fit by the $3/2^{-}$ ($p$ = 0.56) and $3/2^{+}$ ($p$ = 0.75) curves, with the $1/2^{-}$ curve being a poor fit ($p <$ 0.0001).  As the parity is known to be negative, we assign $3/2^{-}$ to this level.  The only acceptable fits to the 4.60 MeV level are from the $3/2^{-}$ ($p$ = 0.09) and $5/2^{+}$ ($p$ = 0.05) curves.  As to the 4.71 MeV level, the best fit is provided by the $5/2^{-}$ curve ($p$ = 0.97), although $7/2^{-}$ ($p = 0.47$) is also reasonable.  In light of our results and the previous tentative assignments, we assign $5/2^{+}$ and $5/2^{-}$ to the 4.60 and 4.71 MeV levels, respectively.     

The 7.17 MeV level is equally well fit by the $11/2^{+}$ and $11/2^{-}$ curves ($p$ = 0.97).  This level was previously assumed to be $(11/2^{-})$\cite{Nes07} simply based on the level scheme of the $^{19}$F mirror nucleus.  We retain both of our possibilities here and constrain this level to be  $J = 11/2$.

\section{Discussion}

The spin-parity assignments from the present work are in general consistent with previous, tentative values.  With regard to estimates of the $^{15}$O($\alpha,\gamma$)$^{19}$Ne rate in XRBs, the assumptions of Ref.\cite{Dav11} for $J^{\pi}$ values of levels between 4.0 and 5.1 MeV are all in agreement with our experimental results.  The assumptions of Ref.\cite{Tan09} differ in the interchange of the assignments to the 4.14 and 4.20 MeV levels; as discussed in Ref.\cite{Dav11}, there had been compelling reasons for either choice.  We note that the quality of the best fits to the 4.60 MeV level has negligible impact on the rate as the $\alpha$ decay branching ratio and mean lifetime have been measured by several groups (see, e.g., Ref.\cite{Dav11}).  With our results, we favor the $^{15}$O($\alpha,\gamma$)$^{19}$Ne rate reported in Ref.\cite{Dav11} over that in Ref.\cite{Tan09}.  The difference between the two recommended rates at the most relevant temperatures of $\approx0.4 - 0.7$ GK\cite{Jos10}, however, is only $\approx30$\%, placing the rate of Ref.\cite{Tan09} well within the rate uncertainty bounds of Ref.\cite{Dav11}.   

We find evidence that the 6.29 MeV state may be either a broad state or a doublet.  This is of interest for the $^{18}$F($p,\alpha$)$^{15}$O rate as if one member of this possible doublet is $J^{\pi}$ = $3/2^{+}$, it will interfere as a subthreshold resonance with other $3/2^{+}$ resonances near the proton threshold, leading to additional uncertainties in the rate due to the unknown sign of the interference\cite{Ade11,Lai13}.  If the state is broad, on the other hand, its measured angular distribution seems inconsistent with $3/2^{+}$\cite{Lai13}.  The lack of any indication of two components from the angular distribution of this structure seems to support the latter option: a broad state that is not $3/2^{+}$.  This is in accord with results from Ref.\cite{Ade11}, in which a single, $3/2^{+}$ level at 6.29 MeV would be expected to have a width $\approx30\times$ smaller than our measured width, based on candidate levels in the mirror nucleus $^{19}$F.  We also note that the good agreement in Fig. \ref{fig3}(a) of the measured angular distributions of the 4034 and 7076 keV states with the theoretical $3/2^{+}$ calculation supports the conclusions in Ref.\cite{Lai13} that none of the members of the triplet of states around 6.4 MeV appears to be $3/2^{+}$, while two subthreshold states around 6.1 MeV may be $3/2^{+}$.  

Finally, our data suggest that one or more new levels may exist near the 6.86 MeV state.  This would not be unexpected given that three as yet unobserved levels within $E_{x}$($^{19}$Ne) = $6.8 - 7.1$ MeV have been predicted based upon comparison with the $^{19}$F mirror nucleus, including two broad states, with widths of 22 and 96 keV\cite{Nes07}.  As well, the non-selective nature of the ($^{3}$He,t) reaction at these beam energies makes this an ideal mechanism by which new states may be observed.  High resolution studies are needed to better determine the nature of the states at 6.29 and 6.86 MeV.  As to the $^{18}$F($p,\alpha$)$^{15}$O rate, the current uncertainties in the structure of $^{19}$Ne around the proton threshold make estimates below $E_{c.m.} = 250$ keV (the lowest energy at which direct measurements exist\cite{Bee11}) unreliable, with uncertainties that are difficult to quantify.

\section{Conclusions}

To improve estimates of the thermonuclear rates of the $^{18}$F($p,\alpha$)$^{15}$O and $^{15}$O($\alpha,\gamma$)$^{19}$Ne reactions at temperatures encountered in classical novae and type I X-ray bursts, respectively, we have measured the $^{19}$F($^{3}$He,t)$^{19}$Ne reaction at 25 MeV using a high-resolution magnetic spectrograph.  We find the levels at $E_{x}$($^{19}$Ne) = 4.14 and 4.20 MeV to be consistent with $9/2^{-}$ and $7/2^{-}$ respectively.  This is in agreement with the tentative assignments in the compilation by Ref.\cite{Til95} and the values assumed for the $^{15}$O($\alpha,\gamma$)$^{19}$Ne rate estimate by Ref.\cite{Dav11}, but not with the values assumed in the rate estimate by Ref.\cite{Tan09}.  We have also measured the $J^{\pi}$ values of four other levels at 4.55, 4.60, 4.71, and 7.17 MeV and we find them to be consistent with previous, tentative assignments\cite{Til95, Nes07}.  A new measurement to both confirm the $\alpha$ decay branching ratios reported in Ref.\cite{Tan07}, as well as to determine the $\alpha$ decay branching ratios of the 4.14 and 4.20 MeV levels separately, would help to improve new estimates of the $^{15}$O($\alpha,\gamma$)$^{19}$Ne rate.  

Our data also indicate the presence of several of the predicted $^{19}$Ne levels near the $^{18}$F+p threshold\cite{Nes07}.  We confirm the triplet of states recently observed around 6.4 MeV\cite{Lai13}, propose that the subthreshold 6.29 MeV state is either a doublet or a broad state, and suggest that the region around the 6.86 MeV state be a target for future high resolution studies to possibly identify additional states.  Indeed, in the absence of suitably intense $^{18}$F beams to extend direct $^{18}$F($p,\alpha$)$^{15}$O measurements\cite{Bar02,Cha06,Ser09, Bee11} to lower energies, we strongly encourage new indirect studies near the $^{18}$F+p threshold to confirm or measure the number of states, their $J^{\pi}$ values\cite{Lai13}, and their proton decay branching ratios\cite{Utk98}.

\section*{Acknowledgments}

It is a pleasure to thank the diligent crew of the MLL tandem accelerator.  This work was supported by the DFG Cluster of Excellence ``Origin and Structure of the Universe".  AP was partially supported by the Spanish MICINN under Grant No. AYA2013-42762.  UK authors were supported by the Science and Technology Facilities Council.

\end{document}